\documentclass{elsart}
\usepackage{graphicx}
\usepackage{amsmath}
\usepackage{amssymb}

\journal{Nuclear Instruments and Methods B}

\newcommand{\Z}{Z_{\rm eff}}

\begin{document}

\begin{frontmatter}
\title{Application of the zero-range potential model to positron annihilation
on molecules}

\author{G. F. Gribakin\corauthref{cor}}
\ead{g.gribakin@qub.ac.uk}
\corauth[cor]{Corresponding author}
\author{C. M. R. Lee}
\ead{c.lee@qub.ac.uk}

\address{Department of Applied Mathematics and Theoretical Physics,
Queen's University Belfast, Belfast BT7 1NN, Northern Ireland, UK}

\begin{abstract}
In this paper we use a zero-range potential (ZRP) method to model positron
interaction with molecules. This allows us to investigate the effect of
molecular vibrations on positron-molecule annihilation using the van der
Waals dimer Kr$_2$ as an example. We also use the ZRP to explore positron
binding to polyatomics and examine the dependence of the binding energy
on the size of the molecule for alkanes. We find that a second bound
state appears for a molecule with ten carbons, similar to recent 
experimental evidence for such a state emerging in alkanes with twelve
carbons.
\end{abstract}

\end{frontmatter}

\section{Introduction} 

The main aim of this work is to achieve better understanding of large
annihilation rates observed for many polyatomic
molecules \cite{Paul:63,Heyland:82,Surko:88,Iwata:95}. In particular, we
use an exactly solvable model to verify the prediction
\cite{Gribakin:00,Gribakin:01} that positron capture into vibrational
Feshbach resonances (VFR) gives rise to a strong enhancement of the
annihilation rate. We also use this model to investigate the dependence
of positron binding energy for polyatomics on the size of the molecule.
Such binding was postulated in \cite{Gribakin:00,Gribakin:01} as a
necessary condition for VFRs.

The annihilation rate for positrons in a gas of atoms or molecules can be
expressed in terms of an effective number of electrons, $\Z$, by
\begin{equation}
\label{annil}
\lambda = \pi {\it r}_0^2cZ_{\rm eff}{\it n},
\end{equation}
where $r_0$ is the classical electron radius, $c$ is the speed of light and
$n$ is the number density of the gas. Measurements by Paul and
Saint-Pierre in the early sixties \cite{Paul:63} indicated unusually large
positron annihilation rates in certain polyatomic molecular gases, with $\Z$
exceeding the actual number of electrons by orders of magnitude. 
They speculated that this might be caused by the formation of
positron-molecule bound states, and later Smith and Paul \cite{Smith:70}
explored the possibility that the large rates might be caused by a vibrational
resonance. Research on the alkanes and similar molecules since
that time \cite{Heyland:82,Surko:88,Iwata:95} uncovered a rapid growth
of $\Z$ with the size of the molecule and very strong chemical sensitivity
of $\Z$. However, only recently a verisimilar physical
picture of this phenomenon has begun to emerge
\cite{Gribakin:00,Gribakin:01,Gribakin:03}. These papers put forward a
mechanism which is operational for molecules with positive positron
affinities, and which involves capture of positrons into VFRs.

Recent measurements of annihilation with a positron beam at high
resolution (25 meV) \cite{Gilb:02,BGS03}, have shown resonances in the energy
dependence of the annihilation rate parameter, $\Z$, of alkane molecules.
Most of the features observed have been unambiguously identified as related to
molecular vibrations. In particular, for all alkanes heavier than methane
$\Z$ displays a prominent C--H stretch vibrational peak. The experiments found
that the magnitude of $\Z$ in the peak increases rapidly with the size of the
molecule (similarly to the increase in $\Z$ observed with thermal
room-temperature positrons \cite{Heyland:82,Surko:88,Iwata:95}). Another
remarkable finding concerns the position of the C--H peak. While for ethane
its energy is close to the mode energy of C--H stretch vibrations ($0.37$ eV),
for heavier alkanes the resonances appear at an energy $\sim 20$ meV lower
for each carbon added. This downward shift provides evidence of positron
binding to molecules. The binding energies observed increase from
about 14 meV in C$_3$H$_8$ to 205 meV in C$_{12}$H$_{26}$. Very recent
experiments show evidence of a second bound state for alkanes with 
12 and 14 carbons \cite{BYS05}.

So far, realistic molecular calculations have not been able to reproduce
enhanced $\Z$. For hydrocarbons and a number of other polyatomics,
calculations have been done using a model positron-molecule correlation
potential in a fixed nuclei approximation \cite{GMO01,OG03}. Such
calculations often provide a reasonable description of low-energy
positron-molecule scattering. However, their results, almost without
exception, underestimate experimental $\Z$, in some cases by an order of
magnitude. This suggests that to describe enhanced
$\Z$, dynamical coupling to molecular vibrations must be included.
Such coupling was considered earlier for diatomics and CO$_2$
\cite{GM99,GM00}, where it had a relatively small effect on $\Z$. (These
molecules most likely do not form bound states with the positron, and do
not possess VFR.) Calculations by the Schwinger multichannel method
\cite{SGL94}, which treats electron-positron correlations {\it ab initio}
for fixed nuclei, also underestimate $\Z$ for molecules such as C$_2$H$_4$
\cite{SGL96} and C$_2$H$_2$ \cite{CVL00} by an order of magnitude \cite{VCL02}.

To examine the effect of vibrations on positron scattering and annihilation,
we consider a simple model of Kr$_2$ dimer using the zero-range potential
(ZRP) method \cite{DO88}. In this model the interaction of the positron with
each of the atoms is parametrised using the atomic value of the scattering
length. It is applicable at low energies when the de Broglie wavelength
of the projectile is much larger than the typical size of the scatterers.
Once ZRP is adopted, the problem of the positron-molecule interaction,
including the vibrational dynamics, can be solved practically exactly. In the
previous paper \cite{Gribakin:02} the interaction between the atoms in the
dimer was treated using the harmonic approximation (HA), which allowed the
vibrational coupling matrix elements to be calculated analytically.
A parabolic potential does not describe well the shallow asymmetric interatomic
potential for a weakly bound van der Waals molecule such as Kr$_2$. In this
work we use the Morse potential to provide a better description of the
molecular interaction. It is a good approximation to the best potential
available for Kr$_2$ \cite{Hal:03}. We examine how the use of a realistic
molecular potential affects the positions and magnitudes of the VFR.
To explore positron binding to
polyatomics we again use the ZRP method. Specifically, we model alkanes
by representing the CH$_2$ and CH$_3$ groups by ZRPs. We investigate the
dependence of the binding energy on the number of monomers and find that a
second bound state emerges for a molecule with ten carbons.

\section{Zero-range model for a molecular dimer}

\begin{figure}[!ht]
\begin{center}
\includegraphics[width=8cm]{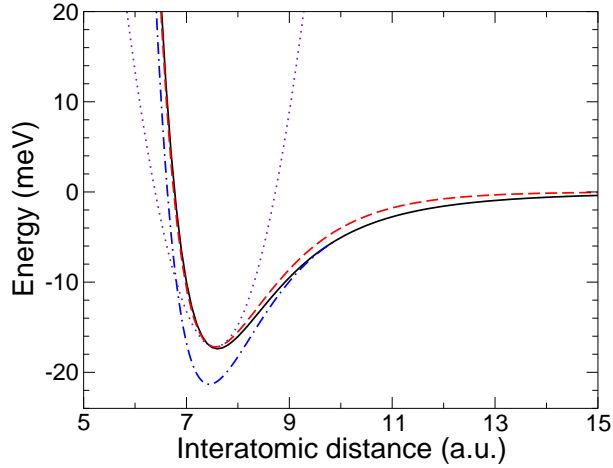}
\end{center}
\caption{Comparison of the best Kr$_2$ potential (solid curve) with
the Morse potential (dashed curve) and harmonic approximation (dotted curve).
Chain curve is the adiabatic potential for the $e^+$Kr$_2$.}
\label{pot_plot}
\end{figure}

In a van der Waals molecule the atoms are far apart and are only weakly
perturbed by each other. This makes it an ideal system for applying the
ZRP method. In this work we model the interaction between the two Kr atoms
using the Morse potential (MP),
\begin{equation} \label{Morse}
U(R) = U_{\rm{min}}[ e^{-2\alpha (R-R_0)}-2e^{-\alpha (R-R_0)}],
\end{equation}
with the parameters $R_0=7.56$ a.u.,
$U_{\rm min} = 6.32\times 10^{-4}~\mbox{a.u.}= 17.2$ meV, and
$\omega = (2U_{\rm min} \alpha ^2/m)^{1/2}= 1.1\times 10^{-4}~\mbox{a.u.}=2.99$
meV \cite{Rad:86,Hub:79}, where $m$ is the reduced mass of Kr$_2$. The energy
eigenvalues and eigenfunctions of the MP
are given by simple analytical formulae \cite{Landau}. In Fig. \ref{pot_plot}
we compare the Morse potential to an accurate fit of the best available
Kr$_2$ potential \cite{Hal:03},
\begin{equation}
\label{PotForm}
V(R)=Ae^{-\alpha R-\beta R^2}-\sum_{n=3}^{8}f_{2n}(R,b)\frac{C_{2n}}{R^{2n}},
\end{equation}
where $\alpha $, $\beta $, and $A$ characterise the short-range part of
the potential, and $C_{2n}$ is a set of six dispersion coefficients. The
function $f_{2n}(R,b)$ is the damping function \cite{Tan:84},
\begin{equation} \label{damp}
f_{2n}(R,b)=1-e^{-bR}\sum_{k=0}^{2n}\frac{(bR)^k}{k!}.	
\end{equation}
The values of the parameters given in \cite{Hal:03} are: 
$\alpha = 1.43158$, $\beta = 0.031743$, $A = 264.552$, $b = 2.80385$,
$C_6 = 144.979$, $C_8 = 3212.89$, $C_{10} = 92633.0$,
$C_{12} = 3.57245\times 10^6$, $C_{14}=1.79665\times 10^8$, and
$C_{16} = 1.14709 \times 10^{10}$ (atomic units are used throughout).

Figure \ref{pot_plot} shows that the Morse potential is close to the best
Kr$_2$ potential, while the HA is valid only in the vicinity of the minimum.
This conclusion is supported by the comparison of the vibrational
spacings. For the MP we have $\omega_{10}\equiv E_1-E_0 = 2.74$ meV,
$\omega_{21} = 2.47$ meV, $\omega_{32} = 2.21$ meV, which agree well
with $\omega_{10}= 2.65$ meV, $\omega_{21} = 2.39$ meV,
$\omega_{32} = 2.12$ meV for the Kr$_2$ potential \cite{Hal:03}. Both
potentials are strongly anharmonic, with $\omega _{n+1,n}$ deviating
markedly from $\omega = 2.99$ meV of HA. Obviously, the MP is
a much better approximation than HA for modelling the Kr$_2$ potential.

In the ZRP model the interaction between a positron and an atom
is expressed as a boundary condition for the positron wavefunction $\psi $,
\begin{equation}\label{eq:zero}
\left. \frac{1}{r\psi}\frac{d(r\psi)}{dr}\right| _{r{\rightarrow}0}=-\kappa_0,
\end{equation}
where $\kappa _0$ is the reciprocal of the scattering length \cite{DO88}.
Positron-Kr scattering calculations yield $\kappa _0=-0.1$ a.u.
\cite{MSC80,DFG96,JL03}.
When applied to a two-centre problem, this condition can be expressed
as
\begin{equation}\label{eq:zero2}
\Psi |_{{\bf r}{\rightarrow }{\bf R}_i}\simeq {\rm const} \times
\left(\frac{1}{|{\bf r}-{\bf R}_i|}-\kappa_0 \right) ,
\end{equation}
where ${\bf r}$ is the positron coordinate, and ${\bf R}_i$ ($i=1,\,2$) are
the coordinates of the two atoms.

Outside the (zero) range of action of the atomic potentials, the positron-dimer
wavefunction can be written as a linear combination of the incident and
scattered waves,
\begin{equation}\label{eq:Psi}
\Psi = e^{i{\bf k}_0 \cdot{\bf r}}\Phi _0({\bf R}) +
\sum_n A_n \Phi _n({\bf R}) \frac{e^{ik_n|{\bf r}-{\bf R}_1|}}
{|{\bf r}-{\bf R}_1|}+
\sum_n B_n \Phi _n({\bf R}) \frac{e^{ik_n|{\bf r}-{\bf R}_2|}}
{|{\bf r}-{\bf R}_2|},
\end{equation}
where ${\bf k}_0$ is the incident positron momentum, $\Phi_n $ is the
$n$th vibrational state of the molecule ($n=0,\,1,\dots $),
and ${\bf R}= {\bf R}_1 - {\bf R}_2$ is the interatomic distance.
Equation (\ref{eq:Psi}) is written for the
case when the positron collides with a ground-state molecule.
The coefficients $A_n$ and $B_n$ determine the excitation amplitude of the
$n$th vibrational state of the molecule, and
$k_n=[k_0^2-2(E_n-E_0)]^{1/2}$ is the corresponding positron momentum.

Applying (\ref{eq:zero2}) gives a set of linear equations for $A_n$ and
$B_n$,
\begin{align}
&(\kappa _0+ik_n)A_n+\sum _m\left(\frac{e^{ik_mR}}{R}\right) _{nm}B_m
= -(e^{i{\bf k}_0\cdot {\bf n}R/2})_{n0}, \label{eq:A} \\
&(\kappa _0+ik_n)B_n+\sum _m\left(\frac{e^{ik_mR}}{R}\right) _{nm}A_m
= -(e^{-i{\bf k}_0\cdot {\bf n}R/2})_{n0}, \label{eq:B}
\end{align}
where ${\bf n}$ is a unit vector along the molecular axis (whose direction
we assume to be fixed during the collision), and the matrix elements are given
by
\begin{equation}\label{n0}
\left(e^{\pm i{\bf k}_0\cdot{\bf n}R/2}\right)_{n0} = \int
\Phi _n^*(R)e^{\pm i{\bf k}_0\cdot {\bf n}R/2}\Phi _0(R)dR ,
\end{equation}
\begin{equation}\label{nm}
\left( \frac{e^{ik_mR}}{R}\right)_{nm} = \int
\Phi_n^*(R)\frac{e^{ik_mR}}{R}\Phi _m(R)dR.
\end{equation}
In HA these matrix elements can be evaluated analytically, (\ref{n0}) --
exactly, and (\ref{nm}) in the leading order \cite{Gribakin:02}. For the
MP we calculated them numerically.

After solving equations ($\ref{eq:A}$)--($\ref{eq:B}$) for $A_n$ and $B_n$,
one finds the total elastic ($0\rightarrow 0$) and vibrational excitation
($0\rightarrow n$, $n=1,\,2,\dots $) cross sections,
\begin{equation} \label{cs_eqn}
\sigma_{0\rightarrow n}=4\pi\frac{k_n}{k_0}|A_n+B_n|^2,
\end{equation}
and the positron annihilation rate,
\begin{equation}
\label{Z_eqn}
Z_{\rm eff}=Z_{\rm eff}^{(0)}\kappa_0^2\sum_ n (|A_n|^2+|B_n|^2),
\end{equation}
where $Z_{\rm eff}^{(0)}$ is the positron-atom $\Z$ at $k=0$ (see
\cite{Gribakin:02} for details). For Kr we use
$Z_{\rm eff}^{(0)}=81.6$ \cite{MSC80}.
Equations (\ref{eq:A})--(\ref{eq:B}) also allow one to determine the energies
of bound states of the positron-dimer system, by looking for the poles of
$A_n$ and $B_n$ at negative projectile energies, i.e., for imaginary
positron momenta $k_0=i|k_0|$.

For doing numerical calculations, the set of equations
(\ref{eq:A})--(\ref{eq:B})
can be truncated by assuming that $A_n=B_n=0$ for $n>N_c$. This means
that only the first $N_c+1$ channels with $n = 0,\,1,\dots ,\,N_c$ are taken
into consideration. At low projectile energies only a small number of channels
are open, and one obtains converged results with a relatively small $N_c$.
In the calculations we used $N_c=15$ and 10, for the HA and MP, respectively.
This value for MP is the total number of bound excited states.

In the single-channel approximation, $N_c = 0$, the HA results practically
coincide with those of the fixed-nuclei approximation, since the matrix
elements (\ref{n0}) and (\ref{nm}) become
$e^{\pm i{\bf k}_0\cdot{\bf n}R_0/2}$ and
$e^{ik_mR_0}/R_0$, respectively (neglecting the 2nd-order and higher
corrections in the small parameter $k_0(m\omega )^{-1/2}$ \cite{Gribakin:02}).
A similar calculation for MP produces slightly different results, because
of the asymmetry of the vibrational ground-state wavefunction, which
gives rise to first-order corrections to these matrix elements.

\section{Results for Kr$_2$}

Table $\ref{tab1}$ shows the values of the bound states (negative) and the
VFRs (positive) of the e$^+$Kr$_2$ complex obtained with MP and in HA. In the
$N_c=0$ approximation the binding energies are $\varepsilon_0 = -3.77$ meV
and $\varepsilon_0=-3.48$ meV for HA and MP, respectively. The binding 
energy for the MP is smaller due to the asymmetry of the potential curve.
The corresponding energies obtained from a multichannel calculation
given in Table \ref{tab1} are lower, because allowing the nuclei to move
leads to stabilisation of the $e^+$Kr$_2$ complex.

\begin{table}[ht!]
\caption{Energies of the bound states and resonances for $e^+$Kr$_2$.}
\begin{tabular}{crr}
\hline
 & \multicolumn{2}{c}{Energy (meV)}\\
\cline{2-3}
Level & HA & MP  \\
\hline	
0 & $-4.23$  &  $-3.89$\\
1 & $-1.41$  &  $-0.74$\\
2 & $ 1.42$  &  $ 2.16$\\
3 & $ 4.25$  &  $ 4.83$\\
\hline
\end{tabular}
\label{tab1}
\end{table}

The ground-state energy of the complex can also be compared to the results
of an adiabatic calculation. For fixed nuclei the energy of the positron
bound state is $-\kappa ^2/2$, where $\kappa $ is a positive root of the
equation $\kappa =\kappa _0+e^{-\kappa R}/R$. Adding this energy
to the Kr$_2$ potential, one obtains the adiabatic potential for the
$e^+$Kr$_2$ complex (chain curve in Fig. \ref{pot_plot}).
Its minimum is about 3.94~meV below that of the Kr$_2$, which is close
to the MP value of $\varepsilon _0$ in table \ref{tab1}.

The first vibrational excitation energy of the $e^+$Kr$_2$ complex,
$\omega' _{10}= \varepsilon _1-\varepsilon _0$, for MP is 3.15~meV, while
in HA it is 2.82~meV. Thus, MP calculations predict a ``stiffening'' of the
vibrational motion of the complex in comparison with that of Kr$_2$.
This effect is caused by a shift of the equilibrium position of the atoms
to the left (see Fig. \ref{pot_plot}), towards the steep repulsive part of
the interatomic potential. Note that in MP, the 2nd bound state with
$\varepsilon_1=-0.74$~meV lies just below the threshold. We will see that
this causes a steep rise in $\sigma_{0\rightarrow 0}$ and $Z_{\rm eff}$ as
$k\rightarrow 0$. In HA this bound state is lower, i.e., further away
from threshold, and its effect on the cross sections is less noticeable.
This combination of a lower binding energy and a greater vibrational
frequency in MP, means that the first resonance observed in the
cross sections and in $\Z$ will be at a greater energy than in HA.

\begin{figure}[!ht]
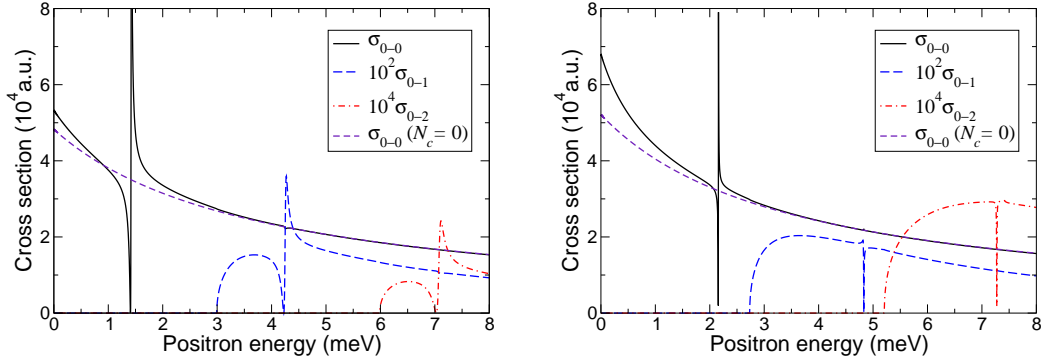

\begin{center}
\begin{minipage}{6.5cm}
\begin{center}
\includegraphics[width=6.5cm]{newHAcs.eps}
\end{center}
\end{minipage}
\hspace{0.5cm}
\begin{minipage}{6.5cm}
\begin{center}
\includegraphics[width=6.5cm]{newMPcs.eps}
\end{center}
\end{minipage}
\caption{Cross sections for positron scattering from Kr$_2$ calculated
using HA (left) and MP (right): solid curve, elastic scattering,
$\sigma_{0\rightarrow 0}$; long-dashed curve, vibrational excitation,
$\sigma_{0\rightarrow 1}$ (times $10^2$); chain curve, vibrational excitation,
$\sigma_{0\rightarrow 2}$ (times $10^4$); short-dashed curve, single-channel
elastic cross section.}
\label{MPHAcs_plot}
\end{center}
\end{figure}

Figure \ref{MPHAcs_plot} shows the elastic and vibrational excitation
cross sections obtained from Eq. (\ref{cs_eqn}) for MP and in HA.
To highlight the effect of resonances, the elastic scattering cross
section has been calculated in both multichannel and single-channel ($N_c=0$)
approximations. The single-channel (``fixed nuclei'') elastic cross
sections from the two calculations are quite close. The multichannel
cross sections are qualitatively similar, the main difference being
in the positions and widths of the resonances and energies of the
vibrational thresholds. The magnitudes of the $\sigma_{0\rightarrow 1}$ and
$\sigma_{0\rightarrow 2}$ cross sections are greater for MP. Another
noticeable difference is in the rise of $\sigma_{0\rightarrow 0}$ towards
zero positron energy in MP calculation.

\noindent 

\begin{figure}[!ht]
\begin{center}
\includegraphics[width=8cm]{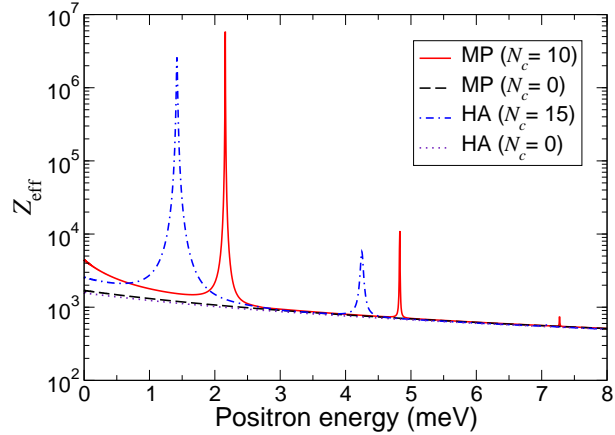}
\caption{$\Z$ for positrons on Kr$_2$ calculated using MP and HA in the
multichannel (solid and chain curves, respectively) and single-channel
(dashed and dotted curves, respectively) approximations.}
\label{ZHAMP_plot}
\end{center}
\end{figure}

Figure \ref{ZHAMP_plot} shows the positron annihilation rate (\ref{Z_eqn})
obtained with and without the coupling to the vibrational motion, i.e.,
from the multichannel and single-channel calculations.
The background (``fixed nuclei'', $N_c=0$) $\Z$ at low positron momenta
is enhanced due to the large positron-Kr$_2$ scattering length.
Such enhancement first predicted in Ref. \cite{Goldanskii:64}, affects
both $\Z$ and the elastic cross section, which in this case are proportional 
to each other, $\Z\sim \sigma _{\rm el}/4\pi $ in atomic units
\cite{Gribakin:00,Gribakin:01}. The effect of VFRs on $\Z$ is much more
prominent than in scattering, with the strongest resonance four orders of
magnitude above the background. The widths of the resonances in MP and HA,
are quite different, e.g., $\Gamma =2.8~\mu$eV (MP) vs. $\Gamma =16.7~\mu$eV
(HA) for the strongest $n=2$ resonance. This difference, also seen in the
scattering cross sections (Fig. \ref{MPHAcs_plot}), means that anharmonicity
of the Kr$_2$ potential reduces the coupling between the incident positron
and vibrationally-excited $e^+$Kr$_2$ compound. A possible explanation for
this is that positron binding has only a small effect on the equilibrium
position of the nuclei (as seen from adiabatic potential curve of $e^+$Kr$_2$
in Fig. \ref{pot_plot}).

\begin{figure}[!ht]
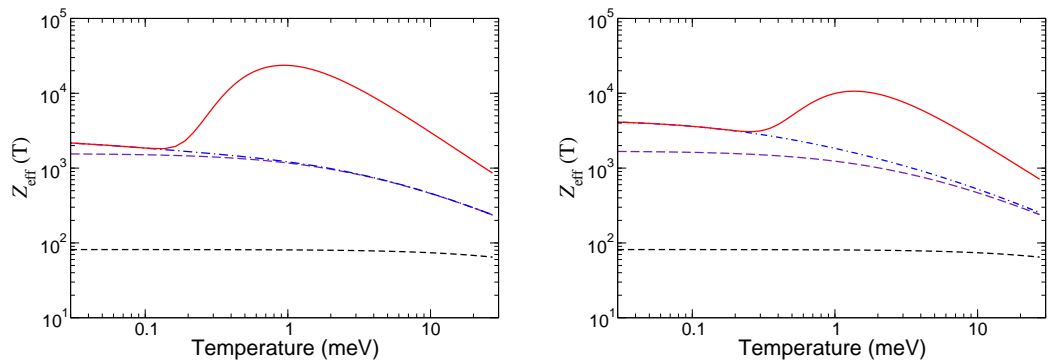

\begin{center}
\begin{minipage}{6.5cm}
\begin{center}
\includegraphics[width=6.5cm]{ZeffnewTHA.eps}
\end{center}
\end{minipage}
\hspace{0.5cm}
\begin{minipage}{6.5cm}
\begin{center}
\includegraphics[width=6.5cm]{ZeffnewTMP.eps}
\end{center}
\end{minipage}
\caption{Thermally averaged $\Z$ for positrons on Kr$_2$, obtained
using HA (left) and MP (right): long-dashed curve, single-channel
approximation; chain curve, multichannel calculation, non-resonant
background; solid curve, multichannel calculation, total, including
the VFR. For comparison, $Z_{\rm eff}$ for Kr is shown (short-dashed curve).}
\label{ZeffT_plot}
\end{center}
\end{figure}

To compare the integral contribution of the resonances, we averaged
$Z_{\rm eff}$ over the Maxwellian positron energy distribution,
\begin{equation}\label{eq:ZeffT}
\bar Z _{\rm eff}(T)=\int_0^{\infty}Z_{\rm eff}(k)
\frac{e^{-k^2/2k_{\rm B}T}}{(2\pi k_{\rm B}T)^{3/2}}4\pi k^2dk.
\end{equation}
Figure \ref{ZeffT_plot} shows $\bar Z_{\rm eff}(T)$ for HA and MP. In both
cases the VFR gives a very large contribution, increasing $\Z$ by an order
of magnitude at $T\sim 1$ meV, i.e., for positron energies close
to that of the resonance. Its effect is seen even at much higher
temperatures, raising $\Z$ above the non-resonant background by a factor
of three for room temperature positrons.

\section{ZRP model of positron binding to polyatomics}

In positron-molecule collisions VFRs occur when the energy of the incident
positron plus the energy released in positron-molecule binding, equals the
energy of a vibrational excitation of the positron-molecule complex. For weakly
bound positron states the latter should be close to a vibrational excitation
energy of the neutral molecule. Hence, by observing VFRs one can obtain
information on the binding energy. This procedure was applied to
electron scattering from (CO$_2$)$_N$ clusters \cite{LBF00}. In this
system the redshift of the VFR was found to increase with the cluster size
by about 12 meV per unit. A simple model of a spherically-symmetric cluster
with a constant potential inside and $-\alpha e^2/r^4$ potential outside
was able to reproduce the dependence of the electron binding energy on the
cluster size.

In a similar way, the measurements of the energy dependence of $\Z$
for alkanes with a high-resolution positron beam allow one to determine
their positron binding energies \cite{Gilb:02,BGS03}. In contrast, an accurate
{\em ab initio} calculation of positron binding to a polyatomic molecule is
probably beyond the capability of present-day theory. Even for atoms,
calculations of positron binding remain a challenging task because of the need
to carefully account for strong electron-positron and many-electron
correlations (see, e.g., \cite{MBR02}).

\begin{figure}[!ht]
\begin{center}
\begin{minipage}{4.5cm}
\includegraphics[width=4.5cm]{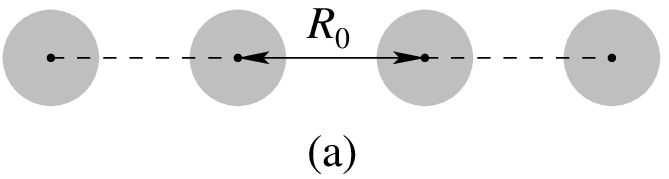}
\end{minipage}
\hspace{0.1cm}
\begin{minipage}{3.8cm}
\includegraphics[width=3.8cm]{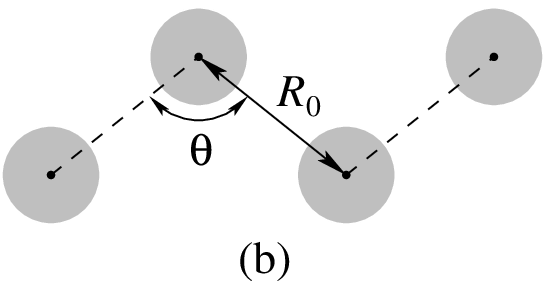}
\end{minipage}
\hspace{0.1cm}
\begin{minipage}{4.7cm}
\includegraphics[width=4.7cm]{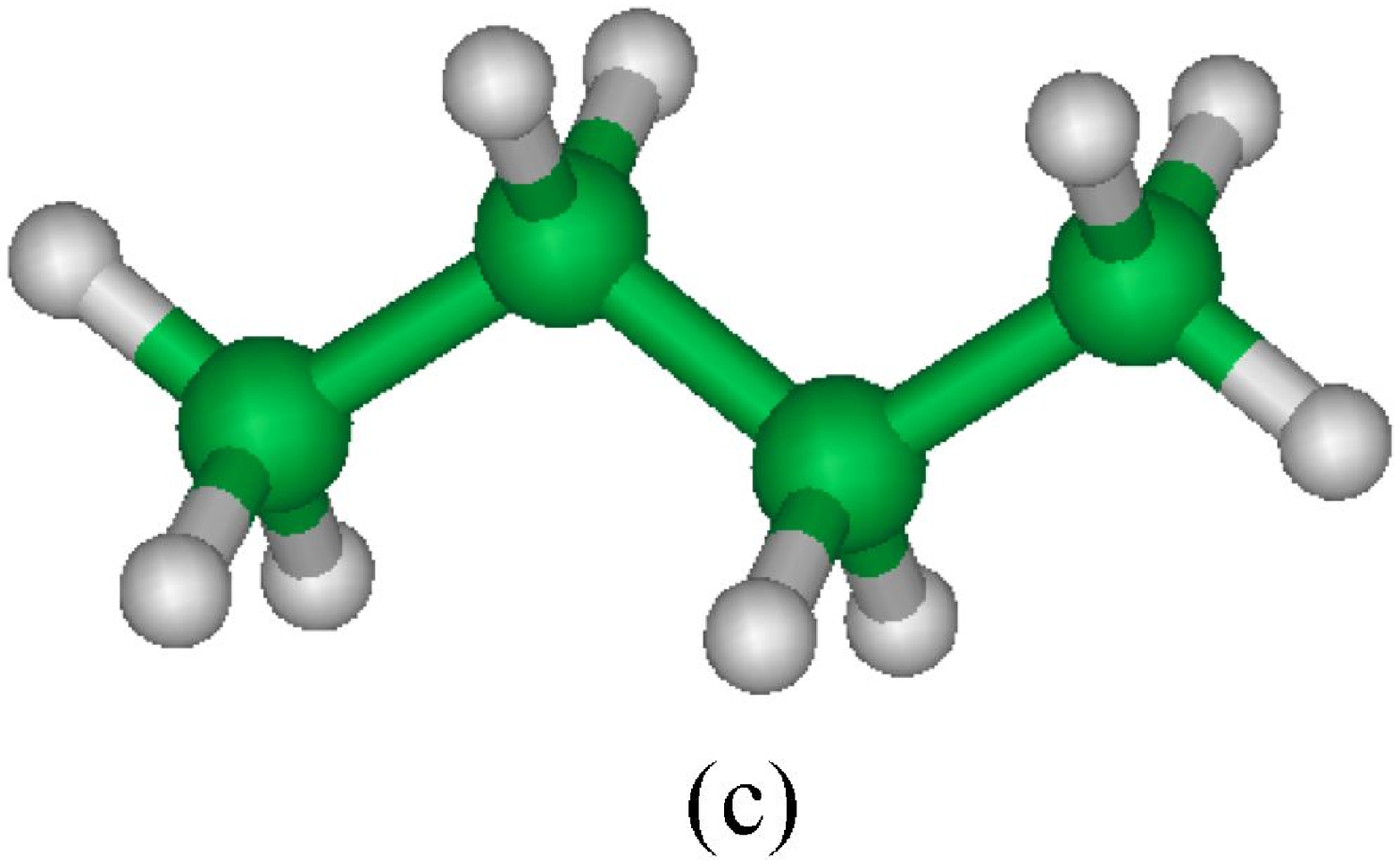}
\end{minipage}
\caption{Zero-range potential models of the alkane molecule (butane,
C$_4$H$_{10}$, is shown). In (a) and (b) the shaded circles represent a
CH$_2$ or CH$_3$ group, while (c) is a true 3D molecular structure \cite{NIST}.
The parameters used are $R_0=2.911$ a.u. and $\theta =113^\circ $.}
\label{alkaneC_plot}
\end{center}
\end{figure}

In this work we have adopted a different approach. To examine positron
binding to alkanes, we model the molecule by representing each CH$_2$ or
CH$_3$ group by a ZRP centred on the corresponding carbon atom. The wave
function of a bound state decreases as $r^{-1}e^{-\kappa r}$ at large
positron-molecule separation $r$, where $\kappa^2/2$ is the binding energy.
For weakly bound states this wavefunction is very diffuse
($\kappa \ll 1$ a.u.), which means that the positron moves mostly
far away from the molecule. The actual binding energy is determined by their
interaction at small distances, and the ZRP model is a simple way of
parametrizing such interaction. It allows us to account for the scaling of the
positron-molecule interaction with the number of monomers (to the extent that
the monomers do not have a large effect on each other), and to use a realistic
geometry of the molecule. We will consider two cases, a straight carbon chain,
Fig. \ref{alkaneC_plot}~(a), and a ``zigzag'' carbon chain,
Fig. \ref{alkaneC_plot}~(b), which matches the actual structure
of the molecule, Fig. \ref{alkaneC_plot}~(c). Unlike the Kr$_2$ model,
the $\kappa_0$ parameter of the ZRP for alkanes is adjusted semiempirically
(see below).

The bound-state positron wavefunction in the field of $N$ ZRP centres
has the form \cite{DO88},
\begin{equation}\label{eq:bound}
\Psi = \sum_{i=1}^N A_i \frac{e^{-\kappa |{\bf r}-{\bf R}_i|}}
{|{\bf r}-{\bf R}_i|}.
\end{equation}
Subjecting it to $N$ boundary conditions (\ref{eq:zero2}) with parameters
$\kappa _{0i}$, we find the positron energy as $\varepsilon _0=-\kappa^2/2$,
where $\kappa $ is a positive root of the equation
\begin{equation}\label{eq:kap}
\det \left[ (\kappa_{0i}-\kappa)\delta_{ij} + \frac{e^{-\kappa R_{ij}}}
{R_{ij}}(1-\delta _{ij})\right] = 0.
\end{equation}
Here $R_{ij}=|{\bf R}_i-{\bf R}_j|$ is the distance between the $i$th and $j$th
ZRP.

For modelling alkanes we choose the distance between the neighbouring ZRPs
equal to the length of the C--C bond. All ZRPs are characterised
by the same value of $\kappa _{0i}=-1.5$. This value ensures that the
molecule with three ZRP centres (which models propane, C$_3$H$_8$) has
a small binding energy, 7 meV for the straight chain, and 12 meV for the
zigzag chain. These values are close to that inferred from experiment
\cite{Gilb:02,BGS03}, where propane is the first molecule for which a
downshift of the C--H peak from the corresponding vibrational energy
can be seen \cite{Gilb:02,BGS03}.

\begin{figure}[!ht]
\begin{center}
\includegraphics[width=8cm]{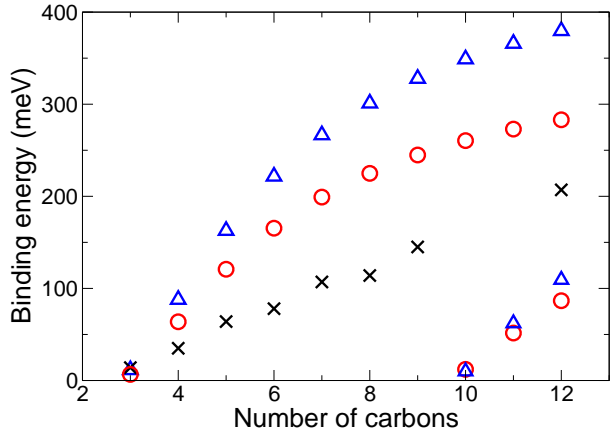}
\end{center}
\caption{Positron binding energies $|\varepsilon _0|$ for alkanes modelled
using straight (circles) and ``zigzag'' (triangles) ZRP chains. Experimental
results (crosses) \cite{Gilb:02,BGS03} are shown for comparison.}
\label{BE4variousC_plot} 
\end{figure}

In Fig. \ref{BE4variousC_plot} the results of our calculations are
compared with the experimental binding energies. As we move from a straight
ZRP chain, Fig. \ref{alkaneC_plot} (a), to a "zigzag" chain,
Fig. \ref{alkaneC_plot} (b), the binding energy increases. This is expected
as the carbon atoms beyond the nearest neighbour become closer to each another.
The overall dependence of the binding
energy on the number of monomers $n$ predicted by the ZRP model is similar
to that of the experimental data, while the absolute values predicted by
our simple theory are within a factor of two of the measurements. One may also
notice that the measured binding energies increase almost linearly with $n$,
while the calculation shows signs of saturation. These discrepancies might
be related to the fact that the ZRP model disregards the long-range
$-\alpha e^2/r^4$ nature of the positron-molecule interaction, which would
restricts its validity to very small binding energies.

A remarkable feature of the model calculations is the emergence of a
second bound state for $n=10$ in both straight and ``zigzag'' chains.
This prediction supports the experimental evidence for the second bound state,
in the form of a C--H peak, which re-emerges at 0.37 eV for dodecane
($n=12$) \cite{Gilb:02,BGS03} and stabilises by about 50 meV for
C$_{14}$H$_{30}$ \cite{BYS05}.

\section{Summary and outlook}

We have used the Morse potential to model the interaction between the atoms
in the Kr$_2$ dimer. We find that the positron binding energies and the
positions and widths of the VFR change compared with the harmonic
approximation. However, the overall picture remains similar, with the lowest
VFR enhancing the Maxwellian-averaged positron annihilation rate by
an order of magnitude to $\Z\sim 10^4$ at $T\sim 1$ meV.

In priniciple, a similar approach could be applied to positron interaction
with other rare-gas dimers and clusters. For Ar and lighter atoms, the
positron-atom attraction is too weak to allow formation of positron bound
states and VFRs. For Xe$_2$, on the contrary, the attraction is much stronger
($\kappa_0\sim -0.01$ \cite{DFG96}). This leads to a much greater
positron-dimer binding energy ($\sim 40$ meV), which means that many
vibrationally excited states of $e^+$Xe$_2$ are bound, and only those with
high $n$ lie above the positron-dimer continuum threshold. The coupling
between these states and the target ground state is extremely weak, and
we have not been able to find any VFR for positrons incident on ground-state
Xe$_2$ in our calculations.

A zero-range potential model for positron binding to alkanes 
yields binding energies that are in qualitative agreement with experiment.
Our calculation predicts the emergence of the second bound state for a
molecule with ten carbon atoms. Such a bound state may have already been
observed for dodecane and tetradecane.

Zero-range potential is an exceptionally simple model. The accuracy of our
predictions is of course limited by the nature of the ZRP model. In particular,
ZRPs disregard the long-range character of the positron-target polarization
attraction. This is a reasonable approximation for very weakly bound states,
but as the binding gets stronger, larger errors may be introduced.
Given the difficulty
of performing {\it ab initio} calculations for positron interaction with
polyatomics, one hopes that more sophisticated yet tractable models
could be developed. They should provide a more accurate description of
positron-molecule binding and capture into vibrational Feshbach resonances.
One may then hope to fully explain the dramatic enhancement of the annihilation
rates and their strong chemical sensitivity for polyatomic molecules.

\section*{Acknowledgements}
We are grateful to L. D. Barnes, C. M. Surko, and J. A. Young for drawing our
attention to the experimental evidence for a second positron bound state
for larger alkanes.

\end{document}